\documentclass[preprint]{aastex}
\usepackage{graphicx}

\newcommand{\HST}{{\it Hubble Space Telescope }}

\begin{document}

\title{A Catalog and Atlas of Cataclysmic Variables - The Living Edition}

\author{Ronald A. Downes}
\affil{Space Telescope Science Institute, Baltimore, MD 21218}
\email{downes@stsci.edu}

\author{Ronald F. Webbink} 
\affil{Department of Astronomy, University of Illinois, 103 Astr. 
 Building, 1002 W. Green St., Urbana, IL 61801}
\email{webbink@astro.uiuc.edu}

\author{Michael M. Shara}
\affil{American Museum of Natural History, Astrophysics Department, 
 Central Park West \& 79th St., New York, NY 10024}
\email{mshara@amnh.org}

\author{Hans Ritter}
\affil{Max-Planck-Institut f\"ur Astrophysik, Karl-Schwarzschild-Str.1, D-85741 
 Garching, Germany}
\email{hsr@mpa-garching.mpg.de}

\author{Ulrich Kolb}
\affil{Department of Physics and Astronomy, The Open University, Walton Hall, 
 Milton Keynes MK7 6AA, United Kingdom }
\email{U.C.Kolb@open.ac.uk}

\and

\author{Hilmar W. Duerbeck}
\affil{WE/OBSS, Free University Brussels (VUB), Pleinlaan 2, B-1050 Brussels,
  Belgium}
\email{hduerbec@vub.ac.be}

\begin{abstract}

The Catalog and Atlas of Cataclysmic Variables (Edition 1 -
\citet{ds93} and Edition 2 - \citet{dws97}) has been a valuable source
of information for the cataclysmic variable (CV) community.  However,
the goal of having a central location for all objects is slowly being
lost as each new edition is generated.  There can also be a long time
delay between new information becoming available on an object and its
publication in the catalog.  To eliminate these concerns, as well as
to make the catalog more accessible, we have created a web site which
will contain a ``living'' edition of the catalog.  We have also added
orbital period information, as well as finding charts for novae, to
the catalog.

\end{abstract}


\section{Introduction}

The Catalog and Atlas of Cataclysmic Variables (Edition 1 -
\citet{ds93} and Edition 2 - \citet{dws97}) has been a valuable source
of information for the cataclysmic variable (CV) community.  One of
the goals of the catalog was to have the basic information on the
objects (i.e. coordinates, type, magnitude range, and finding charts)
in one central location, thus making it easy for observers to obtain
data on the objects.  However, the impracticality of reprinting the
finding charts in their entirety means that, with each new edition,
they are spread among more publications, taking us further from our
goal of a central location.  Furthermore, as new objects are
discovered, and known ones examined in greater detail, the printed
editions cannot keep pace with discovery, A "living" edition is
therefore highly desirable, so that observers can access a complete
and current list of CVs at any time.

For the above reasons, as well as the need to simplify the tracking of
the objects (there are over 1200 objects in the catalog), we have
decided to generate a web-based version of the catalog.  This version
will have all the information (as well some additional information
detailed below) from the first two editions, plus information on over
150 new objects discovered since 1996 May.  Those objects with
revised finding charts will only have one chart presented, thus
eliminating a possible confusion which necessarily exists when
``paper'' catalogs are generated.  The web site will also allow for
easy searching of the catalog, and for generation of basic statistics
(e.g. how many dwarf novae, how many CVs have \HST data, etc.).

The catalog consists of (as of 2000 December) 1034 CVs, and another
194 objects that are non-cvs (objects originally classified
erroneously as CVs). Most of the objects are dwarf novae (40\%), with
another 30\% being novae, and the rest mostly novalike variables.  A
large fraction (90\%) of the CVs have references to published finding
charts, while 64\% of the objects have published spectra (49\%
quiescent spectra and 15\% outburst spectra).

We have taken this opportunity to make several enhancements to the
catalog.  In conjunction with Hans Ritter and Ulrich Kolb, we have
added orbital period data to the catalog; about one-third of the
objects have periods.  The period information is from \citet{rk98},
plus updated and additional values.  In conjunction with Hilmar
Duerbeck \citep{due87}, we now include finding charts of novae (when
possible), and have measured coordinates for many in the \HST GSC v1.1
Guide Star Reference frame (as is the case for the non-novae).
Finally, in the first edition we introduced (out of necessity) a
pseudo-GCVS name for certain objects (e.g. Phe1), which was continued
in the second edition.  With the web-based catalog, these names are no
longer needed, so we will cease generating new ones.  For those
objects that already had such names (some of which have appeared in
subsequent papers in the literature) and now have a formal GCVS
designation, we will adopt the formal GCVS name, although we will keep
the pseudo-GCVS name in the ``Other Name'' field for continuity.

\section{The Site}

The site can be reached via:

http://icarus.stsci.edu/$\sim$downes/cvcat/

and is described in detail below.

\subsection{Home Page}

\begin{figure}
\begin{center}
\includegraphics[width=120mm,angle=0]{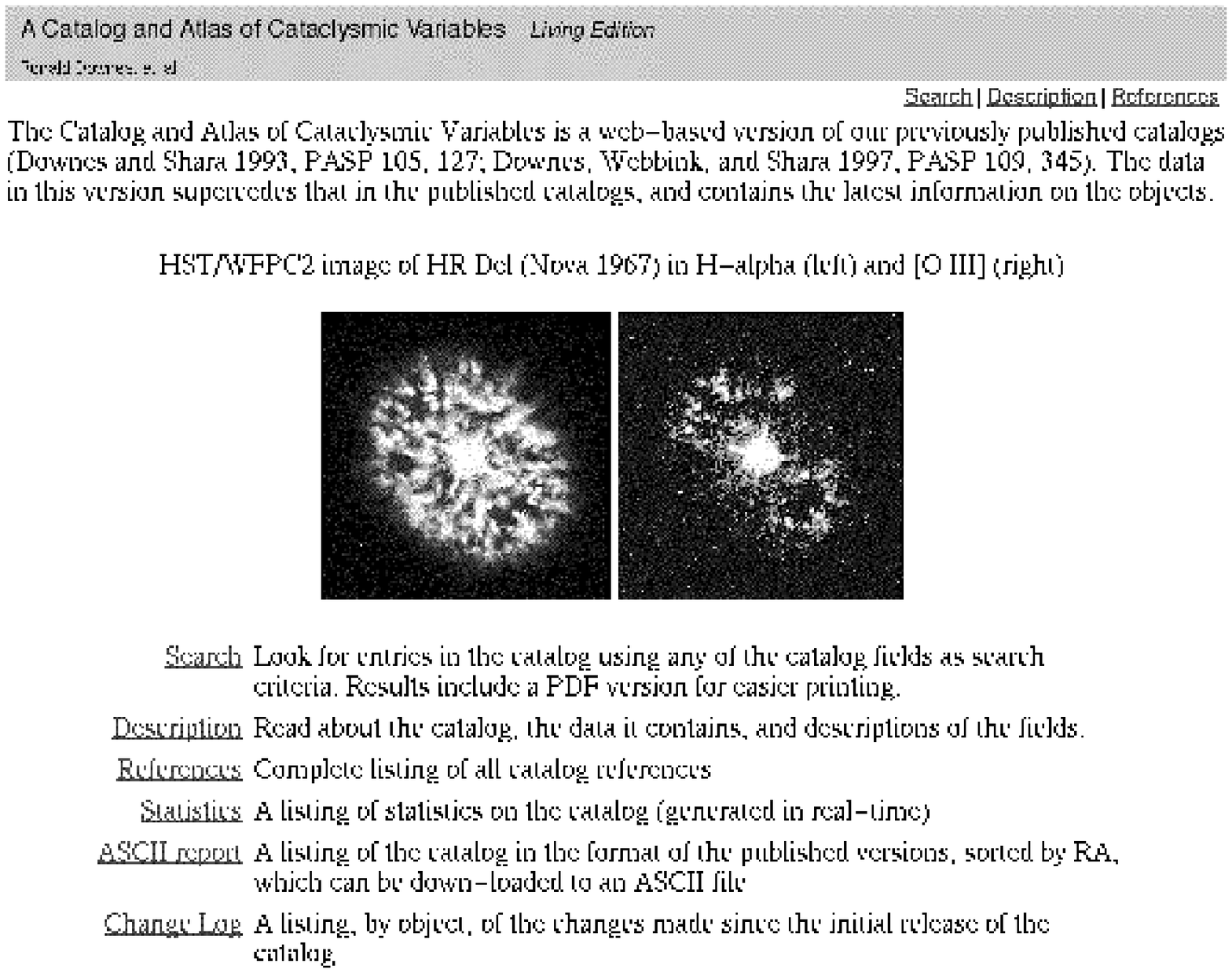}
\vspace*{-1.5cm}
\caption[]{The site Home page.\label{fig1}}
\end{center}
\end{figure}

The Home Page (Figure~\ref{fig1}) for the catalog contains six links:

\begin{itemize}

\item {\bf Search} - a link to the Search Page, from which the catalog
may be accessed.

\item {\bf Description} - a description of the catalog, following the
format of the previous editions. A description of all the fields is
given.

\item {\bf References} - a complete listing of the references
mentioned in the catalog.  Note that from each Individual Object Page,
you can go directly to the reference of interest.

\item {\bf Statistics} - a listing of a fixed set of basic statistics from the
catalog, generated in real-time.

\item {\bf ASCII report} - a listing of the entire catalog in the
format of the previously published versions (i.e. containing most but
not all of the fields), sorted by right ascension.  This output can be
down-loaded to an ASCII file.

\item {\bf Change log} - a listing, by object, of the changes made
since the initial release of this edition

\end{itemize}

\subsection{Search Page}

\begin{figure}
\begin{center}
\includegraphics[width=120mm,angle=0]{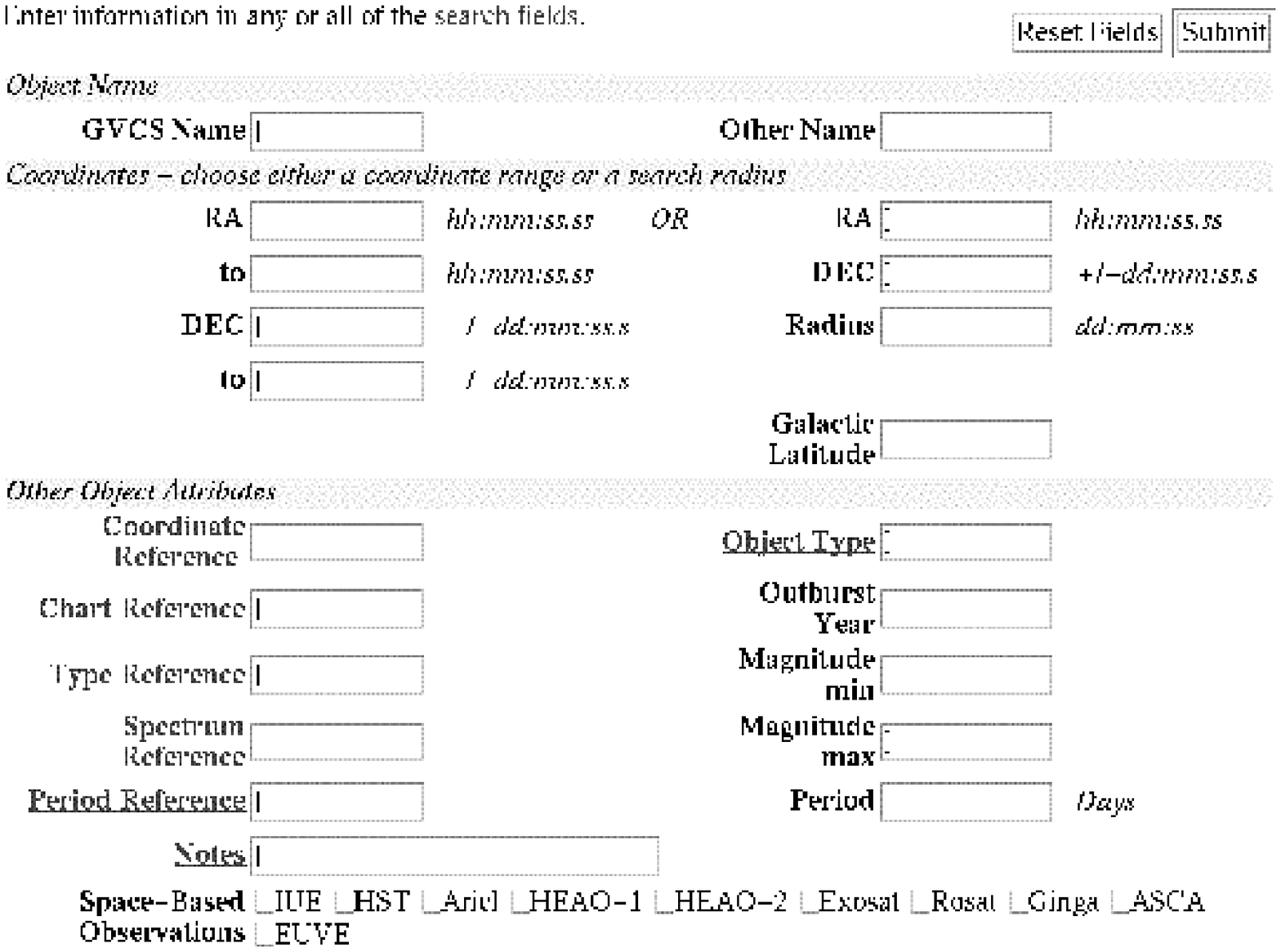}
\vspace*{-.5cm}
\caption[]{The Search page.\label{fig2}}
\end{center}
\end{figure}

The Search Page (Figure~\ref{fig2}) is the main page for access to the
catalog.  It allows the user to search the catalog on any field or
combination of fields. The following text fields can be searched in a
case-insensitive manner: GCVS Name, Other Name, and the five reference
fields (coordinate, chart, type, spectrum, and period); the Object
Type and Notes fields can be searched in a case-sensitive manner. All
textual searches support the use of wildcards.  A coordinate search
may be performed by specifying either a right ascension/declination
range, or by specifying a set of coordinates and a radius.  Numerical
searches (supporting a ``$<$'' and ``$>$'' capability) can be performed for
the following fields: galactic latitude, minimum and maximum
magnitude, outburst year (for novae), and period.  Finally, a search
for space-based observations using any of 10 observatories can be
performed.  An on-line help file is available detailing the search
capabilities for each field, as well as providing instructions for the
use of wildcards.

\subsection{Search Results Page}

\begin{figure}
\begin{center}
\includegraphics[width=120mm,angle=0]{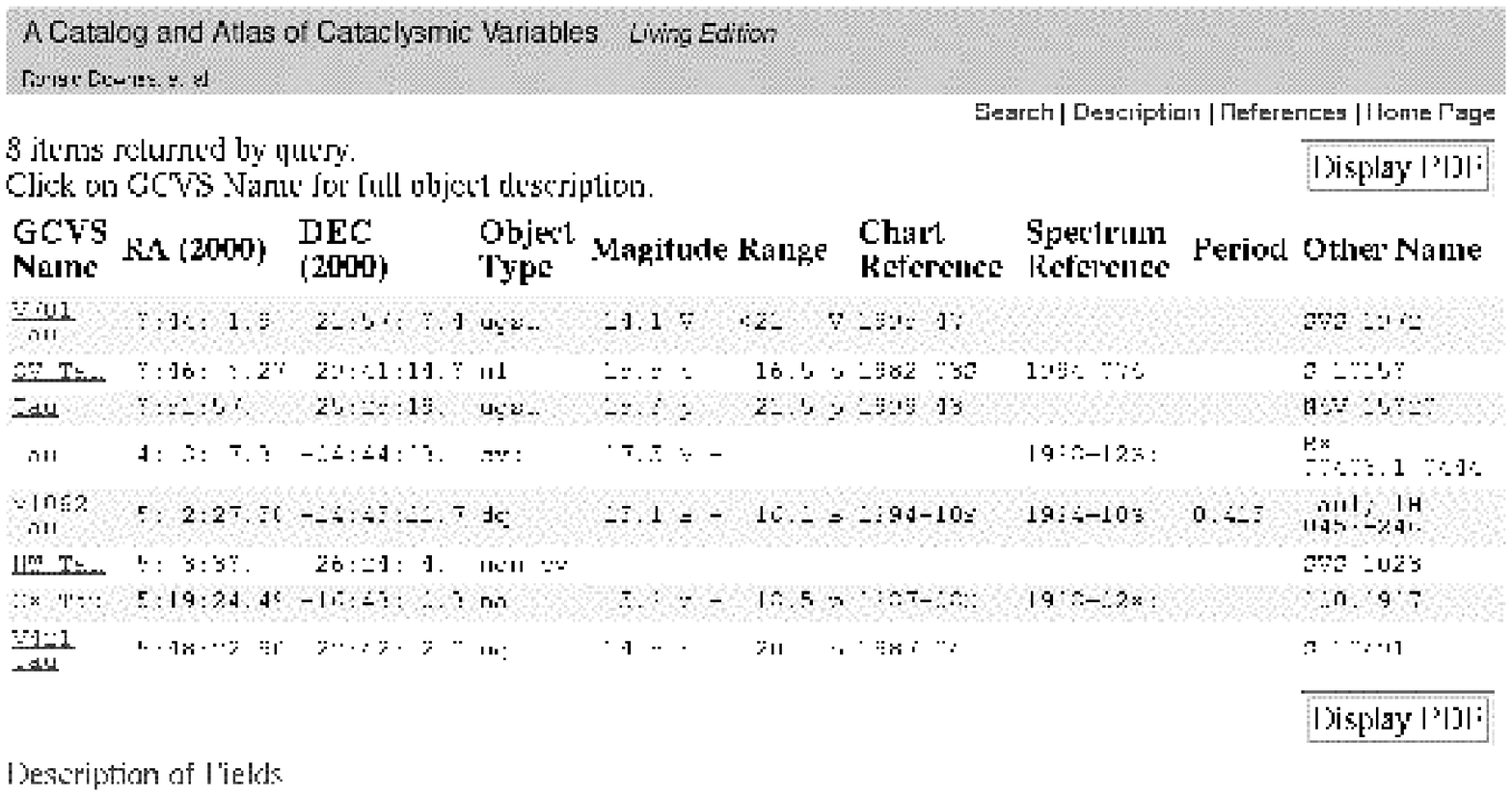}
\caption[]{The Search Results page.\label{fig3}}
\end{center}
\end{figure}

After a search is initiated, the Search Results Page
(Figure~\ref{fig3}) presents the results of the search. This page
indicates the number of objects in the catalog that match the
selection criteria, and presents an abbreviated view of the catalog
entries for such entries, showing the basic information such as the
coordinates, type, magnitude range, and period.  To obtain the full
information (including the finding chart), one clicks on the object of
interest.

\subsection{Individual Object Page}

\begin{figure}
\begin{center}
\includegraphics[width=120mm,angle=0]{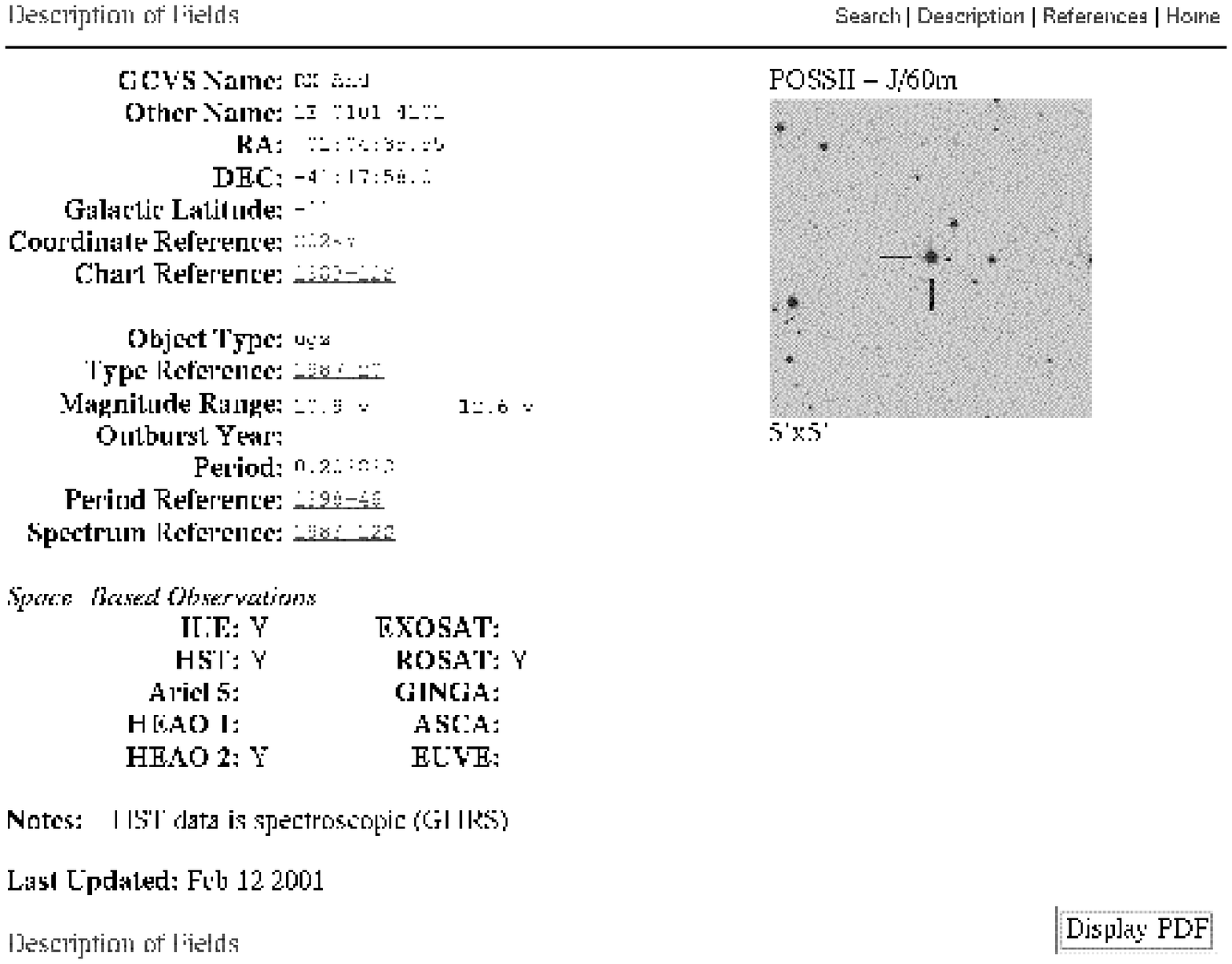}
\caption[]{The Individual Object page.\label{fig4}}
\end{center}
\end{figure}

The Individual Object Page (Figure~\ref{fig4}) presents the complete
information on the selected object.  For the finding charts, the field
size, source (DSS (Digitized Sky Survey), \HST data, ground-based
image), filter/emulsion, and exposure time are given to allow the user
to estimate the depth of the image.  For most objects, the DSS image
is used.  However, for particularly crowded fields (such as in
globular clusters), \HST data is used when available.  Similarly, for
particularly faint targets, ground-based CCD images are provided when
possible.  On this page, one may click on any of the reference codes
to go directly to the full reference on the References page.

\section{Site Maintainance}

We plan to update the site with new objects and information on a
continual basis, although period and spaced-based updates will occur
roughly every six months. We encourage users to inform us of any
updates that should be implemented (e.g. revised identifications, new
objects, etc.), and if appropriate to send us improved/original
finding charts (as either postscript or jpeg images).  The charts will
be particularly useful for recent novae recovered in quiescence, and
for the faintest objects where deep CCD imaging clearly reveals the
correct identification.

\acknowledgments

We wish to thank Anne Gonnella, Steve Hulbert, Calvin Tullos, and Mike
Wiggs for the excellent work in creating the site.  We also wish to
thank Matt McMaster for assistance in generating the multitude of
finding charts, and the Director's Discretionary Research Fund at
STScI for financial support.  Paula Szkody, John Thorstensen, and
Steve Howell provided helpful comments on the initial version of the
site.  RFW gratefully acknowledges the support of NSF grant
AST-9618462, and sabbatical support from STScI.  HWD acknowledges the
hospitality and support of STScI.

\end{document}